\documentstyle[11pt,paspconf]{article}

\def\etal{{\it et~al.\ }}
\def\plotfiddle#1#2#3#4#5#6#7{\centering \leavevmode
\vbox to#2{\rule{0pt}{#2}}
\includegraphics{#1}}

\begin{document}

\title{Rotation Curves and M/L Evolution for Galaxies to $z=0.4$}

\vskip -0.1in

\author{M. A. Bershady$^1$, M. P. Haynes$^2$, R. Giovanelli$^2$,
D. R. Andersen$^3$} \affil{$^1$Department of Astronomy, University of
Wisconsin, 475 N Charter Street, Madison, WI 53706, \\ $^2$Center for
Radiophysics \& Space Research and National Astronomy \& Ionosphere
Center, Cornell University, Ithaca, NY 14853 \\ $^3$Dept. of
Astron. \& Astrophys., Penn State, University Park, PA 16802}

\setlength{\unitlength}{1in}
\begin{picture}(5,0)(0,-3)
\put(-1.0,-0.1)  {\makebox(5,0){To appear in the Proceedings of {\it Galaxy Dynamics} (Rutgers, NJ, 8-12 Aug 1998),}}
\put(-1.15,-0.275){\makebox(5,0){eds. D. R. Merritt, M. Valluri, and J. A. Sellwood (ASP Conference Series)}}
\end{picture}

\begin{abstract}

We present results from an on-going [O~II] $\lambda$3727 and H$\alpha$
rotation-curve survey to study the evolution of the mass-to-light
ratio (M/L) of spiral galaxies selected over a wide, well-defined
range in rest-frame color to $z = 0.4$. Optical--near-infrared
photometry and long-slit spectra are combined to measure residuals
from a fiducial Tully-Fisher (TF) relation as a function of redshift
and galaxy spectral type. While we do find a type-dependence in the
$B$-band TF relation, there is little evidence the dependence has
changed with look-back time. Larger or more detailed kinematic studies
will yield further insights. We demonstrate, for example, how integral
field spectroscopy will provide fundamentally new and more precise
information about the evolution of star-forming disks.

\vskip -0.35in
\end{abstract}


\section{Evolution of star-forming galaxies}

\vskip -0.1in

The study of galaxy evolution has focused largely on the relative
frequency (counts) of photometric properties as a function of redshift
and magnitude, such as luminosity, color, and image structure. To this
it is now possible to add kinematic estimates of mass.  Measurements
of global changes in mass and light, M/L, are sensitive to both the
star-formation and mass assembly histories of galaxies. For some
galaxies, such as starbursts (see Jangren {\it et al.}, these
proceedings, for an extreme example), mass may change abruptly via
merging events. The disks of large galaxies seen today, however, are
presumably long-lived; these dynamically cold disks, if they have
accreted matter in recent times, have done so slowly.  On the other
hand, galaxy formation models require today's disks to form late,
at $z\leq1$ (Mo \etal 1998; Weil \etal 1998). Optical rotation curve
measurements can probe this intermediate redshift regime directly.

Over the past year we have begun a rotation-curve survey at Palomar
Observatory of galaxies selected between $0.2<z<0.4$.  This survey
extends the rotation-curve sample of Bershady, Mihos \& Koo obtained
at Lick Observatory (Bershady 1997) both in redshift, luminosity, and
galaxy type.  While the Lick sample was purposefully limited to the
brightest, late-type galaxies with the reddest optical-infrared
colors, the Palomar sample is representative of the observed
distribution for colors bluer than an un-evolving Sa galaxy and
L$>$0.1L*. The hall-mark of the combined data set is (a) simultaneous,
spatially resolved kinematic measurements made in [O~II] and H$\alpha$
at all redshifts; (b) pre-existing multi-band images, and (c) a well
defined parent sample (Munn \etal 1997) with spectroscopic
redshifts. The combined samples now number 40 galaxies with high-quality
rotation curves. We find [O~II] rotation curves are generally of lower
quality than H$\alpha$ both in spatial extent, S/N, and spectral
resolution. The results presented here use H$\alpha$ rotation curves
only.

\begin{figure}
\plotfiddle{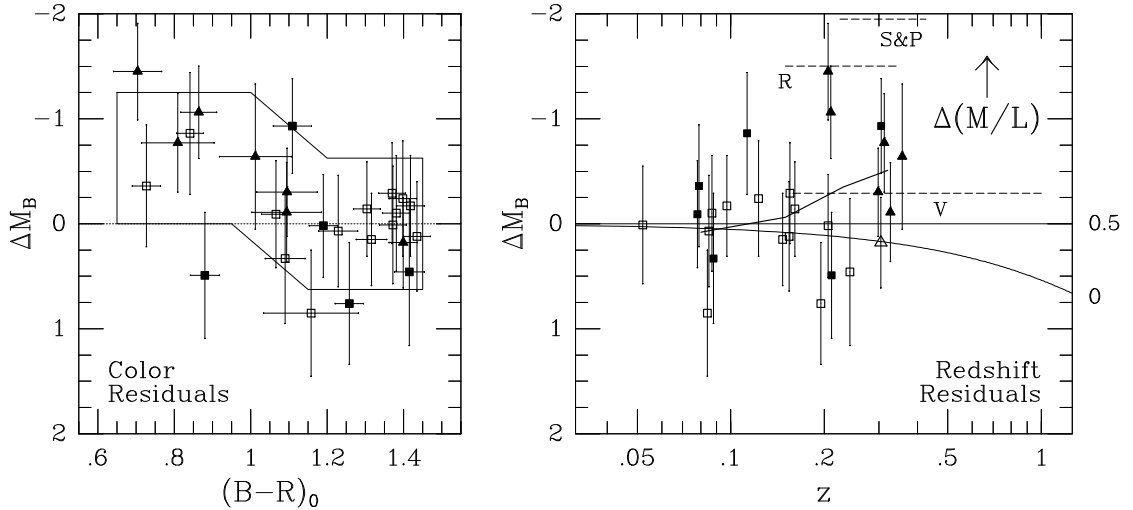}{2.00in}{-90}{60}{60}{-250}{235}

\vskip 0.25in
\caption{\hsize=5.25in \baselineskip 0.165in $B$-magnitude residuals
of the TF relation for our surveys plotted vs rest-frame $B-R$ color
(left panel), and redshift (right panel).  Pierce \& Tully's (1992) TF
calibration, inclination and extinction corrections are adopted;
however, H$_0$=69 km s$^{-1} $Mpc$^{-1}$ (Giovanelli \etal 1997b), and
q$_0$=0.5 are assumed. In both panels our Lick and Palomar samples are
shown as squares and diamonds, respectively.  In the color-residuals
diagram: open symbols $z<0.175$; filled symbols $z>0.175$. The
enclosed area demarks the region occupied by Pierce \& Tully's (1988)
local cluster sample. In the redshift-residuals diagram: Open symbols
$B-R>1.15$; filled symbols $B-R<1.15$.  Average offsets from other
surveys are labeled, dashed lines (V, Vogt \etal 1996, 1997; R, Rix
\etal 1997; and S\&P, Simard \& Pritchet 1998).  Light, solid curves
represent secular changes for different labeled q$_0$. The heavy,
solid curve is the average offset in our combined samples as a
function of redshift, ignoring any color-dependence in the $B$-band TF
relation. However, the observed trend of more negative offsets with
bluer color accounts for the trend in redshift.  We find little
evidence that this color trend evolves with redshift for our sample.}
\vskip -0.17in

\end{figure}

Figure 1 shows the residuals from a fiducial $B$-band Tully-Fisher
relation as a function of rest-frame B-R color and redshift for 26
galaxies currently analyzed from our sample.  Since the cosmological
dependence on magnitude is weak for $z\leq0.5$, the redshift-residuals
diagram can be used to measure evolution: the arrow represents the
effect of decreasing $M/L$. As reported previously (Bershady 1997),
there is significant difference between the residuals we find, and
those found in the surveys of Rix \etal (1997) and Simard \& Pritchet
(1998). Now that our sample contains a wider range of types and
redshifts, we are in a better position to address these
discrepancies. Indeed, we find an offset between blue and red galaxies
within our own sample. Local $B$-band type-dependencies in the
TF-relation are well known (Roberts 1978, Burstein 1982, de
Vaucouleurs \etal 1982, Rubin \etal 1985), and have been measured at
intermediate redshift by Rix \etal (1997). Based on our sample alone,
the type-dependence (as parameterized by color) has not
evolved. Hence, {\it we still find no evidence for evolution of the
TF-relation} to $z = 0.4$.

Assuming this particular class of relatively massive disk galaxies has
evolved at these look-back times, evolution has been both in color,
luminosity, and mass such that galaxy M/L remains within the range
observed locally. Since the range of observed rotation speeds at a given
luminosity and color does not change with redshift in our sample, we
have no compelling evidence that mass has changed substantially. A
likely possibility is that the relative fraction of galaxies (red to blue)
changes, but at a fixed rotation speed and size (or mass). To confirm
this, larger intermediate-$z$ samples are needed for detailed comparison to
local templates (e.g. Giovanelli \etal 1997a). To refine inferences of
mass evolution, independent measures of luminosity evolution are also
needed. Promising diagnostics include surface-brightness, rotational
asymmetry (e.g. Conselice 1997), and emission line-strength compared,
for example, at fixed color and line-width.

\vskip -0.3in
\null

\section{Disk Kinematics and Integral Field Spectroscopy}

\vskip -0.1in

In addition to gathering larger kinematic samples with more
sophisticated photometric diagnostics, there is yet another direction
for intermediate redshift kinematic surveys: spatial maps of disk
velocity and velocity-dispersion. These can be readily obtained from
Fabry-Perot images or integral-field spectra of optical and
near-infrared emission lines. Detailed kinematic studies are
invaluable for several reasons. For example, modeling suggests that
when distant galaxies are observed with a single slit, slit
mis-alignment and spatial resolution significantly affect the inferred
rotation velocities.  While some of these systematics can be corrected
reliably with spatially-resolved spectra (Vogt \etal 1996, Bershady \&
Mihos 1999), the corrections assume axisymmetric velocity
fields. Velocity maps remove many of these problems, most notably
kinematic vs photometric position-angle mis-alignments, and
inclination errors due to disk asymmetries. Figure 2 illustrates the
potential power of integral-field spectroscopy for mapping velocity
fields in galaxies; with 10m-class telescopes, these measurements can
be made easily at intermediate redshift (e.g see Andersen \&
Bershady, these proceedings). Detailed kinematic studies will be
necessary to calibrate larger, statistical surveys using only a single
fiber or slit per target.

Non-axisymmetric components of disk velocity fields are interesting in
their own right, particularly if the incidence (or amplitude) of the
asymmetries change with look-back time. Such an increase might be
expected if the 'harassment' or mass-accretion rate were higher or
more irregular in the past. Ideally, comparisons would be made between
kinematic and photometric asymmetry (e.g. Kornreich \etal 1998),
using optical or near-infrared light for both measures.

Most challenging will to measure stellar velocity dispersions in
galaxy disks, near and far. Such measurements are of critical
importance for estimates of disk mass and M/L (see Bosma, Fuchs, and
Quillen, these proceedings). While Fabry-Perot's are powerful for
single emission-line measurements over a large dynamic range in angle,
for absorption line measurements of single targets, integral-field
spectrographs with large \'etendue will be the instruments of choice.

\vskip -0.17in

\acknowledgments We thank C. Mihos \& D. C. Koo for allowing us
to present data in advance of publication.

\begin{figure}
\plotfiddle{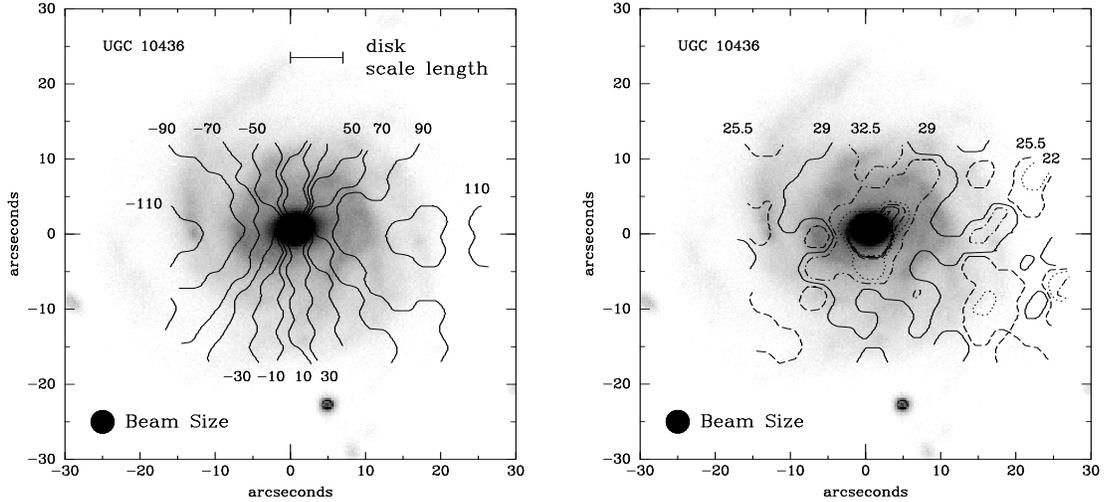}{2in}{0}{85}{85}{-265}{-285}

\vskip 0.35in
\caption{\hsize=5.25in \baselineskip 0.165in R-band image of UGC 10436
at $z\sim0.03$ superimposed with contours of velocity ({\bf left
panel}) and velocity dispersion ({\bf right panel}), as measured in
H$\alpha$ (km s$^{-1}$) with WIYN's Densepak (Andersen \& Bershady
1999). The photometric axis ratio implies an inclination of $<$
10$^{\circ}$), while modeling of the kinematic map yields an
inclination of $\sim$35$^{\circ}$, indicating a photometric asymmetry
in the disk. Velocity dispersions ($\sigma$) have been corrected for
beam-smearing; the effect is small outside of the central beam. These
high signal-to-noise H$\alpha$ measurements can be obtained with
integral field spectroscopy in 0.5 hours on a 3.5m telescope. With a
few times higher \'etendue, comparable measurements could be made in
stellar absorption in reasonable exposure times.}
\vskip -0.5in
\end{figure}

\null

\newpage

\title{Mass Estimates of Starbursting Galaxies: Line Widths versus
Near-IR Luminosities}

\author{A.~Jangren$^1$, M.~A.~Bershady$^{1,2}$, C.~Gronwall$^3$}


\affil{$^1$Department of Astronomy \& Astrophysics, Penn State
University, University Park, PA 16802 
\\ $^2$Dept. of Astronomy, University of Wisconsin, Madison, WI 53706
\\ $^3$Astronomy Department, Wesleyan University, Middletown, CT 06459}


\keywords{}

\bigskip

{\bf Introduction.}  We present preliminary estimates of the stellar
masses of 6 compact, narrow emission-line galaxies (CNELGs) at
redshifts $0.10 < z < 0.35$, drawn from earlier studies (Koo {\it et al.}
1994, 1995, Guzm\'an {\it et al.} 1996).  In previous work, Guzm\'an
{\it et al.} (1996, 1998) use galaxy sizes, [O III] $\lambda 5007$
emission line-widths $\sigma$, and the shape of the light profiles to
estimate dynamical galaxy masses, $M_{\sigma}$, of the CNELGs. The
small line-widths ($\sim$50 km/s) and sizes (R$_e \sim$ 2.5
h$_{50}^{-1}$ kpc) yield virial masses of order
$1-5\cdot10^9\,M_{\odot}$.  However, Guzm\'an {\it et al.} point out
that the total masses are likely to be up to $\sim$4 times larger: the
underlying stellar population may be twice as big as the galaxy
half-light radius $R_e$ which was used to estimate the galaxy size,
and the [O III] line-widths may underestimate the internal
velocities. A key assumption is that the line-widths reflect
virialized motion, rather than turbulent motion due to stellar winds.
Here, we make estimates of galaxy stellar masses based on $K$ band
luminositites, and compare these masses to those derived from emission
line-widths and optical sizes.

\medskip

{\bf Analysis and Results.} $K$ band images of the CNELGs were
obtained with the KPNO Mayall 4-m telescope in 1994. To derive galaxy
stellar masses from near-IR luminosities, the simplest case is to
assume a constant $(M/L)_K$ for all objects -- a useful approximation
for normal galaxies, where the $K$ band flux mainly comes from the old
stars (Rix 1993). For starbursting galaxies the situation is more
complex: both old and young stars contribute substantially to the
near-IR flux (Leitherer \& Heckman 1995), making it difficult to
assign a mass-to-light ratio to the galaxy without detailed knowledge
of the burst age and strength (Figure 1, left panel).

Instead, we use population synthesis models ($\sim0.25 Z_{\odot}$, and
no reddening) to estimate the overall $(M/L)_K$ for a starburst
superimposed on an underlying, older stellar population (Schmidt {\it
et al.} 1995, Worthey 1994, Leitherer \& Heckman 1995). From $(B-V)$
and $(V-K)$ rest-frame colors we then estimate the ages of starbursts
of different strengths; the burst characteristics determine the $(M/
L)_K$ ratio of the galaxy. Based on the individual $(M/L)_K$ of each
object and its $K$ band luminosity, the stellar masses, $M_{NIR}$,
were estimated. However, the adopted metallicity and reddening values
of the burst model, if incorrect, can lead to significant systematic
effects: underestimating $Z$ or reddening will lead us to assume
older, weaker bursts with higher $(M/L)_K$ ratios, but will weakly
affect our estimate of the $K$ band luminosity. In this scenario,
$M_{NIR}$ is an upper limit to the true stellar mass. The precise
value of $(M/L)_K$ as determined from broad-band colors can also vary
depending on the detailed burst properties and star-formation history.
 
Nonetheless, the resulting stellar masses are correlated with the
line-width masses: on average, the stellar masses are 4 -- 10 times
larger (Figure 1, right panel). This is roughly the expected result;
the line-width method systematically underestimates the dynamical
galaxy masses, and the starburst models ignore the effects of dust
reddening and may lead us to overestimate the stellar masses. Implicit
in this comparison is that dark matter does not dominate the potential
on the small physical scales of the starburst region. While both the
line-width and the $K$ band methods for deriving galaxy mass are
problematic, used in concert they can place useful limit on stellar
$(M/L)$ and masses.

\medskip

{\bf Acknowledgements.} Research was funded by NASA from
STScI/AR-07519, STScI/GO-07875 and NAG5-6043.

\setcounter{figure}{0}

\begin{figure}

\includegraphics{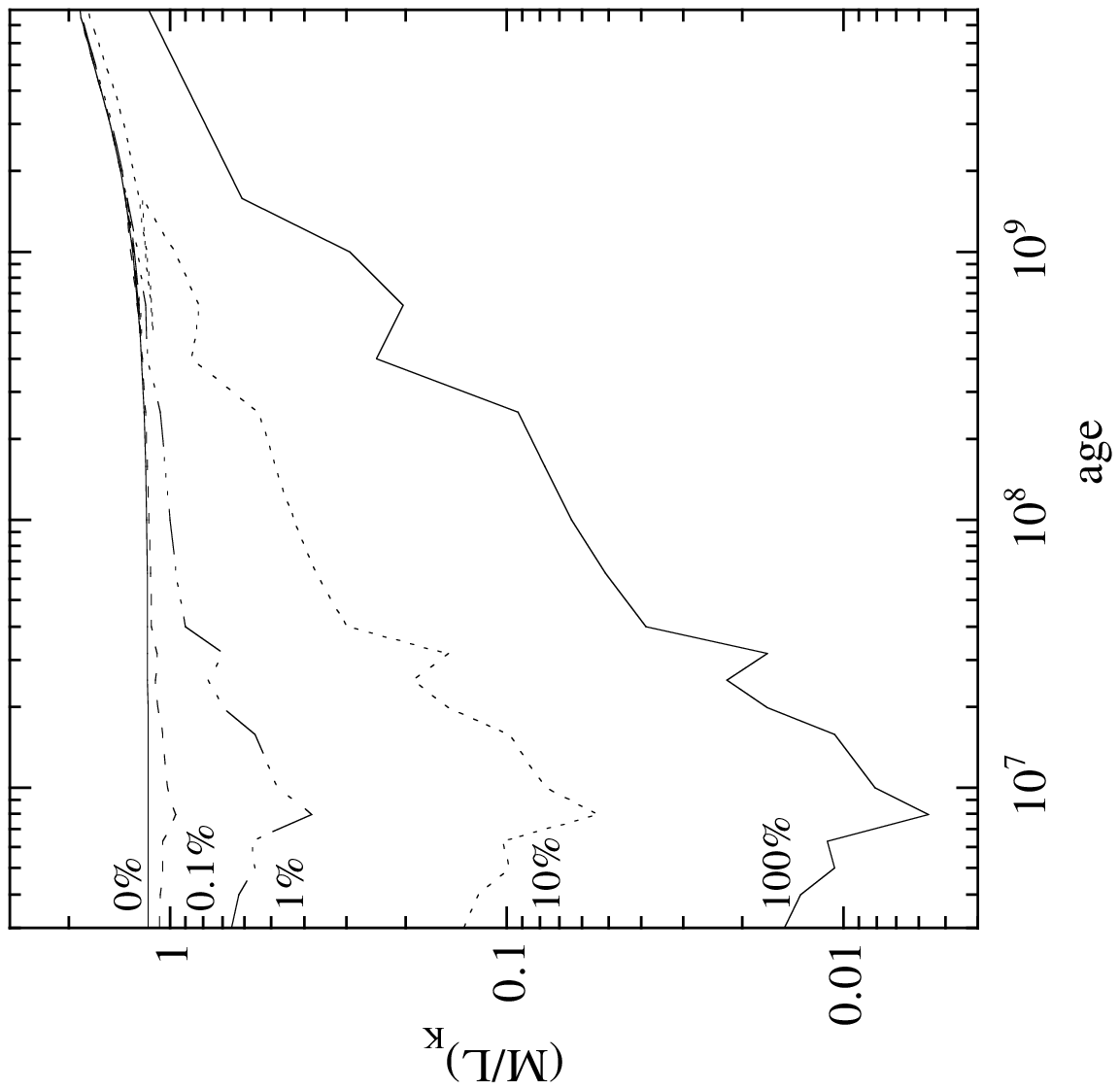}

\includegraphics{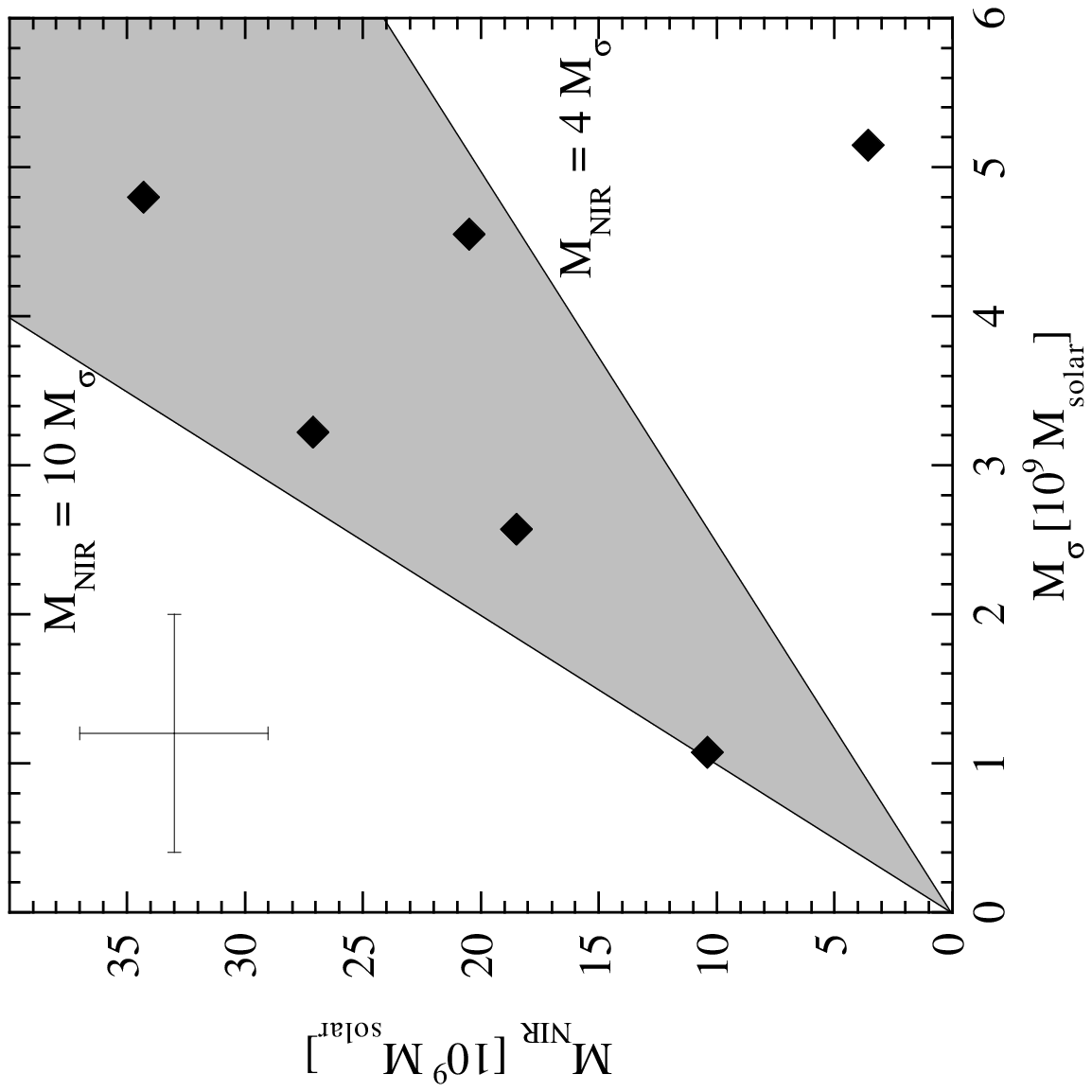}

\vspace{2.7in}
\caption{{\bf Left panel:} The $K$ band mass-to-light ratio,
$(M/L)_K$, is plotted versus the age of the starburst for bursts of
varying strengths (100\% -- pure burst, 0\% -- older stellar
population). {\bf Right panel:} The masses derived from the near-IR
luminosities ($M_{NIR}$) are plotted versus the emission line-width
masses ($M_{\sigma}$). Typical error bars are shown. $M_{NIR}$ is
typically 4 -- 10 times larger than $M_{\sigma}$ (shaded area).}
\vskip -0.1in
\label{fig-1}
\end{figure}

\medskip

\noindent{\bf References}

\noindent Guzm\'an,~R., Jangren,~A., Koo,~D.~C., Bershady,~M.~A.,
Simard,~L. 1998, \apj, 

495, L13

\noindent Guzm\'an,~R., \etal 1996, \apj, 460, L5

\noindent Koo,~D.~C. \etal 1995, \apj, 440, L9

\noindent Koo,~D.~C. \etal 1994, \apj, 427, L9

\noindent Leitherer,~C., \& Heckman,~T.~M. 1995, \apjsupp, 96, 9

\noindent Rix,~H.-W. 1993, \pasp, 105, 999

\noindent Schmidt,~A.~A., Alloin,~D., Bica,~E. 1995, \mnras, 273, 945

\noindent Worthey,~G. 1994, \apjsupp, 95, 107

\newpage

\title{Galaxy Kinematics with Integral Field Spectroscopy}
\author{David R. Andersen}
\affil{Department of Astronomy and Astrophysics, Penn State University,
University Park, PA 16802}
\author{Matthew A. Bershady}
\affil{Department of Astronomy, University of Wisconsin -- Madison,
Madison, WI 53706}

\section{Science Mission}

The integral field unit (IFU), nicknamed ``Spider'' (Figure 1), is a
fiber optic array under construction for the 9m Hobby-Eberly
Telescope's (HET) Medium Resolution Spectrograph (MRS). This
contribution updates previous reports (Bershady et al. 1998), and
represents the final design for this instrument.  Spider is optimized
for integral field spectroscopy for kinematic studies of nearby and
moderately distant galaxies, but will have more general application to
moderate spectral resolution studies of extended sources at low
surface-brightness (Bershady 1997). The IFU will be capable of
delivering simultaneous rotation curves and disk velocity dispersions
over a range of look-back times.

Rotation curve and disk velocity dispersion measurements from integral
field spectroscopy allow for separate estimates of the disk and total
masses and allow us to probe the axisymmetry of spiral disks.  A
critical limitation has been the inefficiency of traditional
long--slit measurements made on 4m~class telescopes.  Integral field
spectroscopy using high--throughput echelle spectrographs promise to
open a new window on the mass and mass distributions of galaxies over
a range of look--back times.

\section{Performance}
The IFU covers an area of 15 by 15 arcseconds using 1 arcsecond (200
$\mu$m) fibers to densely sample slits at 4 position angles at
resolutions from 5500 to 14500.  Our 200 $\mu$m array will be
installed during the commissioning phase of the MRS (mid-1999).  This
fiber-fed echelle spectrograph has a resolution of 10,900 for a 1
arcsec aperture, and initial spectral coverage from 0.5-0.95 $\mu$m in
a red beam. The IFU coupled to the MRS spectrograph is expected to
reach a limiting surface-brightness in V of 22 at S/N = 10 per
spectral resolution element per fiber at R = 10,900 in one hour
(Figure 2); this assumes a peak throughput of 15\% for the HET plus
MRS system (Ramsey 1995).  The large telescope aperture and fiber size
of Spider yields large \'etendue (50 m$^2$ arcsec$^2$). We are unaware
of other fiber arrays planned or in existence which have comparable
\'etendue and spectral resolution. Consequently, this IFU fills a
niche for moderate resolution spectroscopy at low surface-brightness.

\newpage

\setcounter{figure}{0}

\begin{figure}[h]
\vbox to 2.8in{\rule{0pt}{5.5in}}
\includegraphics{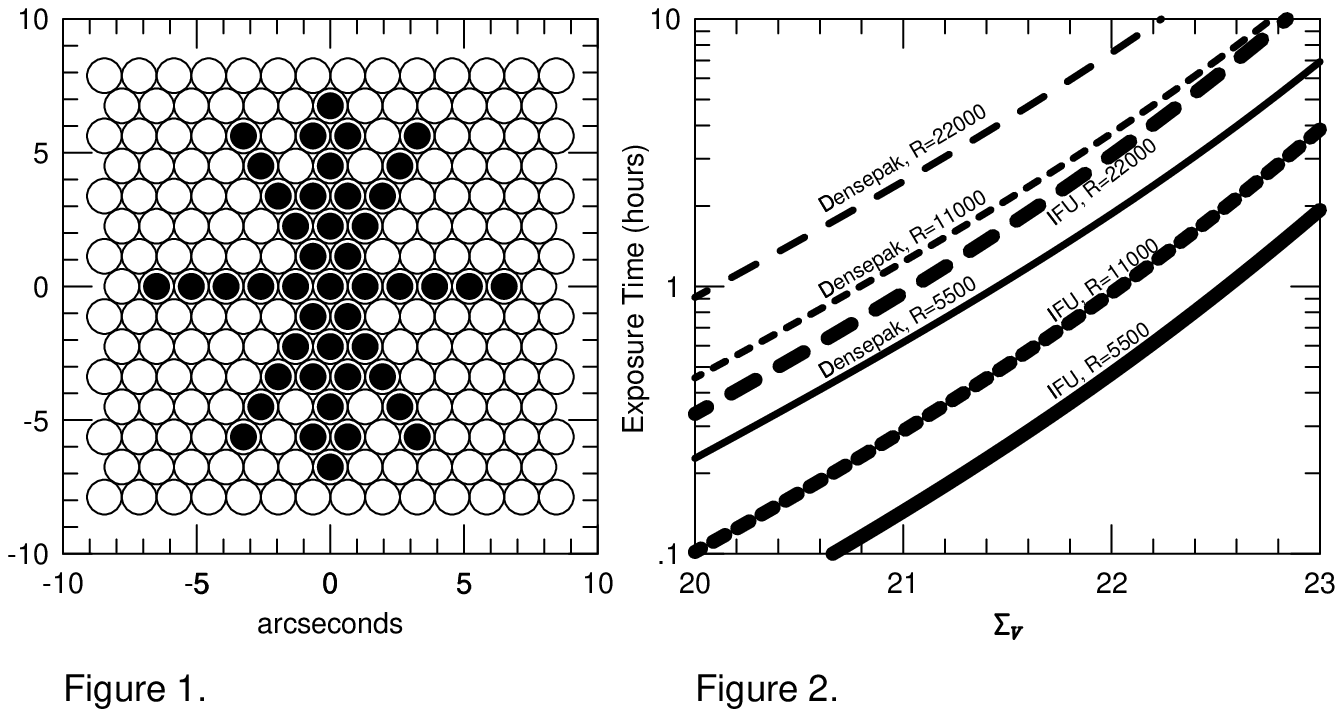}

\caption{Focal plane design of Spider integral field unit to be
installed during the commissioning phase for the HET MRS (mid-1999).
The fiber cores are 200 $\mu$m (1~arcsecond) in diameter with a 20:3:3
core to clad to buffer ratio. {\bf Dark fibers} (45 fibers) are
science fibers that go through to the spectrograph; {\bf open fibers}
are short (2.5 cm) packing fibers that maintain mechanical
rigidity. Six sky fibers (not shown) will be mounted on an
independently positionable probe. }

\end{figure}

\begin{figure}[h]
\vbox to -.4in{\rule{0pt}{5.5in}}

\caption{Surface Brightness (Johnson $V$, assuming sky is $\Sigma_V =
21.9$ arcsecond$^{-2}$) versus exposure time needed in order to reach
a signal to noise of 10 per resolution element for Spider (1 arcsecond
fibers on the 9m~HET) and the DensePak fiber array (3 arcsecond fibers
on the 3.5m~WIYN telescope; Barden \& Wade 1988) at resolutions of
5500, 11000, and 22000. {\bf The IFU will be able to reach $\Sigma_V =
22$ arcsecond$^{-2}$ at resolution of 11,000 in one hour.}}

\end{figure}

\acknowledgments

This instrument is supported by NSF/AST 96-18849. We are indebted to Sam
Barden for consultation and advice on fabrication.

\end{document}